\documentclass[prb,reprint,showpacs,amsmath,amssymb]{revtex4-1} 
\usepackage{bm}
\usepackage{graphicx}
\usepackage{color}

\begin{document}

\title{Thermal boundary conductance across rough interfaces probed by molecular dynamics}

\date{\today}

\author{Samy Merabia}\email{samy.merabia@univ-lyon1.fr}
\affiliation{Institut Lumi\`{e}re Mati\`{e}re, UMR5306 Universit\'{e} Lyon 1-CNRS, Universit\'{e} de Lyon 69622 Villeurbanne cedex, France}

\author{Konstantinos Termentzidis}
\affiliation{LEMTA-UMR7563, CNRS and University of Lorraine, F-54506 Vandoeuvre les Nancy, France}

\begin{abstract}
In this article, we report the influence of the interfacial roughness on the thermal boundary conductance between two crystals, using molecular dynamics.
We show evidence of a transition between two regimes, depending on the interfacial roughness: when the roughness is small, the boundary conductance is constant taking values close to the conductance of the corresponding planar interface. When the roughness is larger, the conductance becomes larger than the planar interface conductance, and the relative increase is found to be close to the increase of the interfacial area. The cross-plane conductivity of a superlattice with rough interfaces is found to increase in a comparable amount, suggesting that heat transport in superlattices is mainly controlled by the boundary conductance. These observations are interpreted using the wave characteristics of the energy carriers.
We characterize also the effect of the angle of the asperities, and find that the boundary conductance displayed by interfaces having steep slopes may become important if the lateral period characterizing the interfacial profile is large enough. Finally, we consider the effect of the shape of the interfaces, and show that the sinusoidal interface displays the highest conductance, because of its large true interfacial area. All these considerations are relevant to the optimization of nanoscale interfacial energy transport. 	

\end{abstract}

\pacs{68.35.Ja, 07.05.Tp, 44.10.+i}

\maketitle
\section{Introduction}
The existence of a finite thermal boundary resistance between two solids has important practical consequences, especially in the transport properties of nanostructured materials. When the distance between interfaces becomes submicronic, heat transfer is mainly controlled by the interfacial phonon transmission, which in turn governs the thermal boundary resistance. In certain applications, such as electro-optical modulators~\cite{schneider1990}, optical switching devices\cite{wagner2007} and pressure sensors~\cite{robert1999}, a low resistance is desired to favor energy flow. In thermoelectric devices on the contrary, a large resistance is preferable so as to generate large barriers for a wide class of phonons~\cite{wan2010,hashibon2011,qiu2011}.
Two strategies may be followed in order to tune the value of the boundary resistance between two solids. Either, the solid/solid interaction is changed through the coupling with a third body, which is usually a self-assembled monolayer~\cite{losego2012,obrien2013}. The other possibility is to modulate the interfacial roughness~\cite{gotsmann2013}. This latter direction has been illustrated experimentally through chemical etching~\cite{hopkins2010,hopkins2011,duda2012}. However, a theoretical model describing the effect of the interfacial roughness on the thermal boundary conductance at room temperature is still lacking~\cite{kechrakos1990,kechrakos1991}. Note that the role of the interfacial roughness on the Kapitza conductance has been underlined a long time ago, in the context of liquid Helium/metal interfaces at very low-cryogenic temperatures~\cite{amrit2010,shiren1981}. 
At these temperatures, the phonon coherence length may be comparable with the typical heights of the interface, leading to strong phonon scattering which is put forward to explain the high values of the conductance experimentally reported, as compared with the classical acoustic mismatch theory which assumes planar interfaces~\cite{adamenko1971}.
Such considerations have received less attention for room temperatures solids, probably because in this case the phonon coherence length is very small. 

Understanding the role of the interfacial roughness has also important consequences in the transport properties of superlattices, which are good candidates for thermoelectric conversion materials, thanks to their low thermal conductivity~\cite{tritt2004}. Designing superlattices with rough interfaces has been recently achieved, opening an avenue for reducing the thermal conductivity in the direction perpendicular to the interfaces~\cite{termentzidis2011}. Again, the physical mechanisms at play in the heat transport properties of rough superlattices have not been elucidated so far. Molecular dynamics offers a privileged route to understand the interaction between the energy carriers in a solid and the asperities of the interface~\cite{termentzidis2011_2,rajabpour2011,termentzidis2009}.  
In this article, we use molecular dynamics to probe interfacial heat transfer across model rough interfaces. Because of the difficulty to determine the temperature jump across a non-planar interface, we have used transient simulations which enable to compute the thermal boundary resistance characterizing rough interfaces. 
In section~\ref{simulations}, we describe the structures used to probe the conductance of rough interfaces. 
In section~\ref{GK}, we explain the methodology retained to extract the thermal boundary conductance from molecular dynamics. The simulation results are presented and discussed in section \ref{results}. We first concentrate on model interfaces made of isosceles triangles.

For these model interfaces, we present the results for the thermal conductance as a function of the interfacial roughness and 
interpret the results using a simple acoustic model in subsection \ref{results_roughness}. In the following section, we characterize the effect of the angles of the asperities. Finally, in sub-section~\ref{results_shape}, we have appraised the effect of the interfacial shape. We discuss the consequences of this work in the Conclusion.

\section{Structures and sample preparation \label{simulations}}

\begin{figure}
\includegraphics[width=0.9\linewidth]{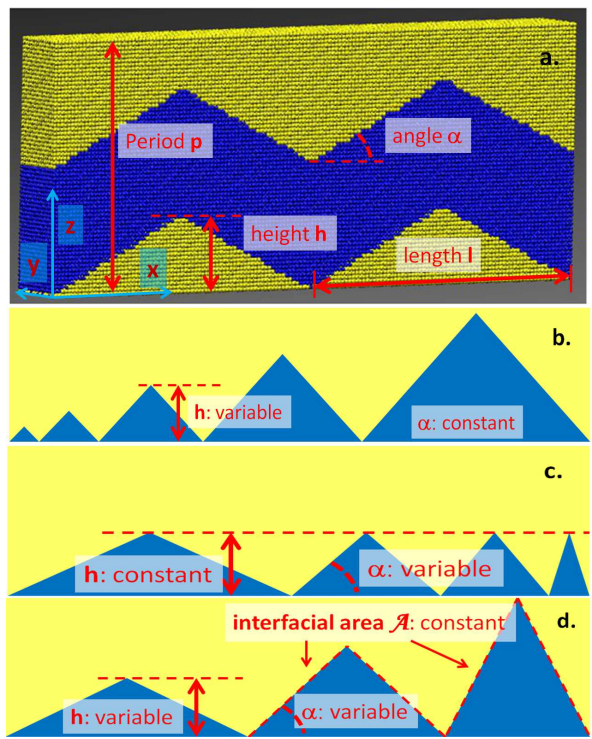}
\caption{(Color online) Diagram showing the different parameters that will be varied, for the triangular shaped interfaces.
Schematic representation of the different parameters that have been varied: 
Top middle (Fig. 3b): interfacial height at a constant value of the angle $\alpha=45 \deg$; Medium (fig 3c): angle $\alpha$ at a constant value of the interface height $h$. Bottom 
(Fig 3d): angle $\alpha$ and interfacial height $h$ at a constant value of the interfacial area $\mathcal A$.}
\label{schematic_parameters}
\end{figure}

We will consider model rough interfaces, constructed from two perfect fcc Lennard-Jones solids whose interface is orientated along the crystallographic [100] direction, as represented in figure~\ref{schematic_parameters}. We introduce some 2D roughness in the $xz$ plane, where $x$ and $z$ denote respectively the [100] and [001] directions. As we use periodic boundary conditions in all spatial directions, the system studied is similar to a superlattice. 
The dimension in the $y$ direction has been fixed to $10$ $a_0$, where $a_0$ is the lattice parameter, while the dimension in the $z$ direction -the superlattice period-has been varied between  $5$ and $40$ $a_0$. \newline
All the atoms of the system interact through a Lennard-Jones potential $V_{\rm LJ}(r) = 4\epsilon \left((\sigma/r)^{12}-(\sigma/r)^6\right)$ truncated at a distance $2.5 \sigma$. A single set of energy $\epsilon$ and diameter $\sigma$ characterizes the interatomic interaction potential. As a result, the two solids have the same lattice constant $a_0$.
To introduce an acoustic mismatch between the two solids, we have considered a difference between the masses of the atoms of the two solids, characterized by the mass ratio $m_r =m_2/m_1$. In all the following, we will use $m_r=2$, which has been shown to give an impedance ratio typical of the interface between Si and Ge~\cite{termentzidis2009}. From now on, we will use real units where $\epsilon=1.67 \; 10^{-21}$ J; $\sigma =3.4 \; 10^{-10}$ m and $m_1=6.63 \;10^{-26}$ kg, where these different values have been chosen to represent solid Argon. With this choice of units, the unit of time is $\tau=\sqrt{m\sigma^2/\epsilon}=2.14$ ps, and the unit of interfacial conductance is $G = k_B/(\tau \sigma^2) \simeq 56 MW/K/m^2$.  
The different interfaces have been prepared as follows: first the structures have been generated by mapping the space with fcc structures using the lattice parameter of the fcc LJ solid at zero temperature~\cite{chantrenne2003}: $a_0(T=0 K)= 1.5496 \sigma$. The structures have been equilibrated at the two final finite temperatures $T=40$ K and $T=18$ K using a combination of a Berendsen, a Nos\'e Hoover thermostat and a barostat at $0$ atm~\cite{frenkelsmit}. The total equilibration time lasts one million time steps which corresponds to a total time of $4,28$  ns. The equilibrium lattice parameters have been found to be: $a=1.579 \sigma$ at T=$40$ K and $a=1.5563 \sigma$ at T=$18$ K. 

In this article, we considered different types of rough interfaces, as represented in Fig.~\ref{schematic_parameters}.  The first one consists of triangular shaped interfaces, having a constant angle $\alpha=45\deg$ with respect to the $xy$ plane and a variable height $h$. In the second type of interface analyzed, we keep constant the interfacial height $h$ and we vary the angle $\alpha$. In the third case, both the angle and the height are varied keeping constant the interfacial area $A$. Finally the effect of the shape of the interfaces has been also appraised considering totally rough interfaces, small triangles juxtaposed on triangular interfaces, square and wavy shaped interfaces. 
This analysis covers all the possible parameters that might be involved in the geometry of the interfaces with the aim to quantify their influence on phonon interfacial transport.

\section{Thermal boundary conductance from transient simulations \label{GK}}
In this section, we briefly review the methodology adopted to probe the interfacial conductance between two solids, using transient out-of-equilibrium simulations. 

\begin{figure}
\includegraphics[width=0.9\linewidth]{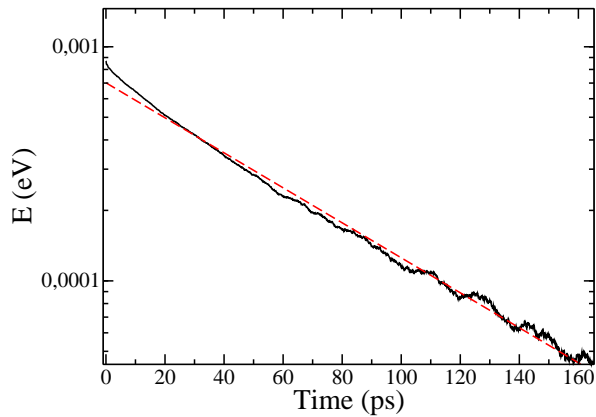}
\caption{(Color online) Energy decay of the heated solid obtained using transient non-equilibrium simulations. Dashed lines show the exponential fit. 
The parameters are: Total Length=$40$ $a_0$; temperature $T=40$ K; mass ratio $m_r=2$.}
\label{energy_decay}
\end{figure}

Generally speaking, there are different methods to extract the boundary conductance from molecular simulations: either a net heat flux $q$ is applied through the coupling of two energy reservoirs, and one measures the finite temperature jump $\Delta T$ across the interface~\cite{stevens2007,landry2009}. This allows to compute the interfacial conductance $G_K = q/\Delta T$. For the rough interfaces that we will consider in the following, it may be difficult to clearly identify a temperature jump, especially if the roughness is large. On the other hand, transient non-equilibrium simulations do not require to resolve spatially the temperature field in the vicinity of the interface, and for this reason are well adapted to the determination of the conductance of imperfect interfaces.
The principle is akin to the thermoreflectance technique, and consists in heating instantaneously one of the two solids and record the temporal evolution of the temperature of the hot solid~\cite{shenogin2004,hu2011}. The conductance $G_K$ is then obtained from the time $\tau$ characterizing the thermal relaxation of the hot solid:
\begin{equation}
\label{transient}
G_K = \frac{3 N_1 k_B}{4 A \tau} 
\end{equation}
where $N_1$ is the number of atoms of solid $1$, $k_B$ is the Boltzmann's constant, and $A$ is the interfacial area.
Because the temperature of the heated solid may display some oscillations which may make the determination of the time constant $\tau$ difficult, we have rather used the decay of the energy $E_1$ to extract $\tau$:
\begin{equation}
\label{mechanical_energy}
E_1^m= \sum_{j \in 1} \frac{1}{2}m \vec v_j^2 + \sum_{j,k \in 1} V(\vec r_j- \vec r_k)
\end{equation}
where the second term corresponds to the interatomic potential, here supposed to be pair-wise.
An example of the time decay of the energy during thermal relaxation of the hot solid is displayed in Fig.~\ref{energy_decay}, showing that the exponential decay hypothesis is reasonable.

%
In practice after equilibration of the system, we have heated one of the two solids by an amount of $18$ K and followed the thermal relaxation of the hot solid during a time interval after which the energy has decreased by a factor $10$.
To remove any contribution stemming from internal phonon scattering in the heated solid, we have run in parallel simulations across the interface between identical solids, and calculated the corresponding internal resistance. 
The Kapitza conductance calculated in this article has been obtained after having substracted this internal resistance: 
\begin{equation}
\label{interfacial_conductance}
1/G_K = 1/G_{12} - 1/G_{11}
\end{equation}
where $G_{12}$ is the conductance measured for the interface between solid $1$ and solid $2$; and $G_{11}$ is the conductance measured 
between identical solids using eq.~\ref{transient}.
This procedure has been followed for all the systems studied in this article. Finally, 
for the simulations discussed in this article,
we have used between $5$ and $10$ independent configurations depending on the system size, to determine the value of the Kapitza conductance, and the error bar has been found to be typically $15$ percents.  

\section{Results \label{results}}
In this section, we present the simulation results obtained using the transient simulations, as detailed above. We will successively study the effect of the interfacial roughness, the angle of the asperities, and the shape of the interface. A summary of the different parameters that will be varied is depicted in Fig.~\ref{schematic_parameters} and Fig.~\ref{illustration_shape}.

\subsection{Effect of the superlattice period and number of periods~\label{finite_size_effects}}
\begin{figure}
\includegraphics[width=1.2\linewidth]{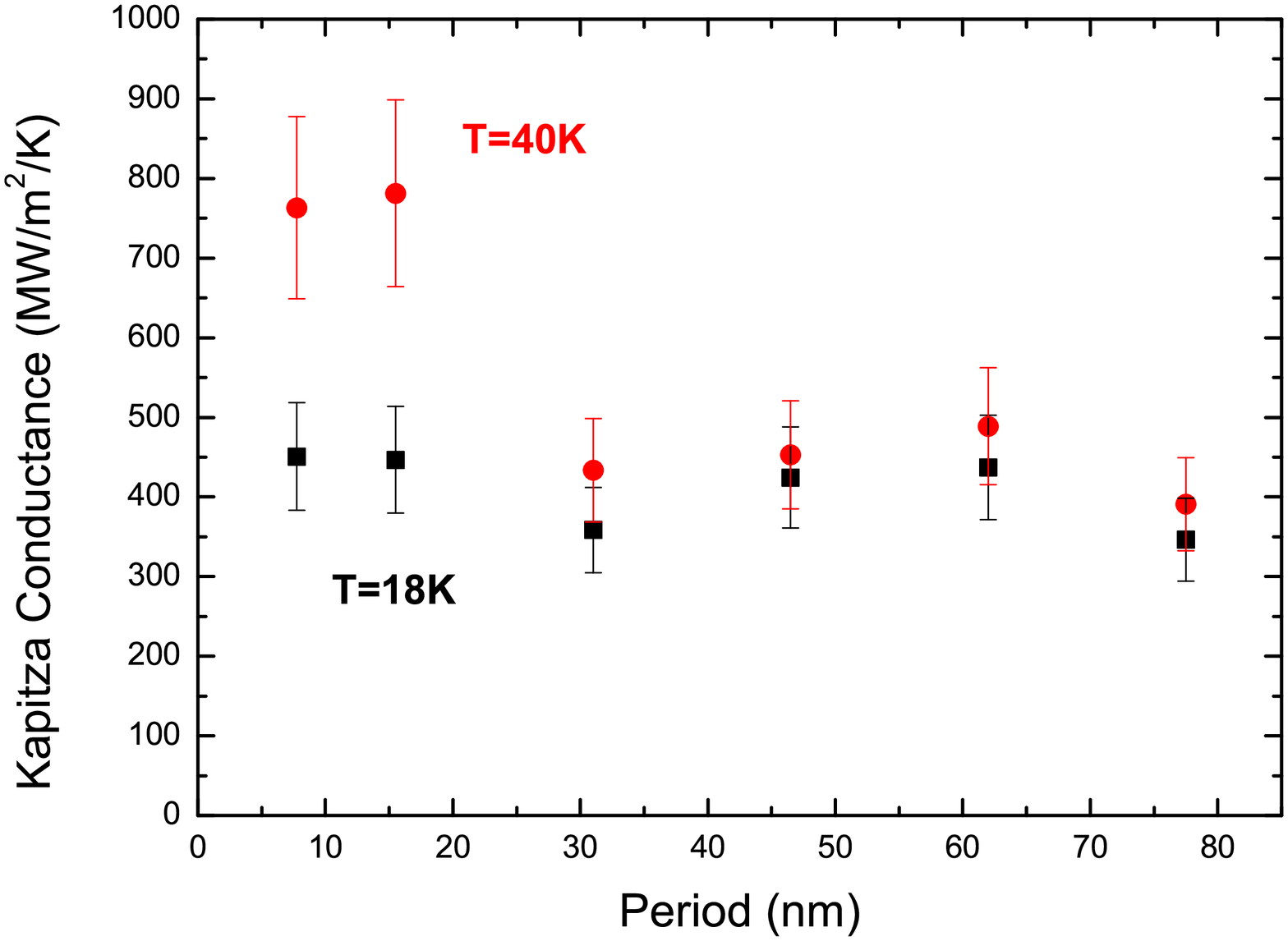}
\caption{(Color online) Thermal boundary conductance of isosceles triangular shaped interfaces having a roughness height $h=32$ MLs, as a function of the period $p$ defined in fig~\ref{schematic_parameters}. Top: $T=40$ K; Bottom: $T=18$ K. Lines are guides to the eye. 
The mass ratio is $m_r=2$.}
\label{fig_finite_size_effects}
\end{figure}

In this section, we first quantify finite size effects in the determination of the conductance of rough interfaces,
as measured by eq.~\ref{transient}.
It is important to note that the system simulated is not a single isolated interface, but rather a superlattice because of the periodic boundary conditions. It is thus relevant to assess the influence of the superlattice period on the thermal conductance as measured by eq.~\ref{transient}.
To this end, we will consider model rough interfaces, made of isosceles triangles, as depicted in fig.~\ref{schematic_parameters}.
In figure \ref{fig_finite_size_effects}, we report the conductance of triangular shaped interfaces having a fixed roughness height  
$32$ MLs, and a varying period. 
For the two temperatures considered, the thermal conductance is found to decrease with the system size for short periods, and then saturates for periods larger than $30$ nm. The increase of the conductance for thin layers may be explained because long wavelength phonons will not see two independent interfaces but rather a single one.
A similar trend has been also reported in lattice dynamics simulations~\cite{zhao2005} and Green function calculations~\cite{zhang2007}. For thick layers, the conductance measured is constant,
and has converged to the value characterizing an infinitely thick film.

\begin{figure}
\includegraphics[width=0.8\linewidth]{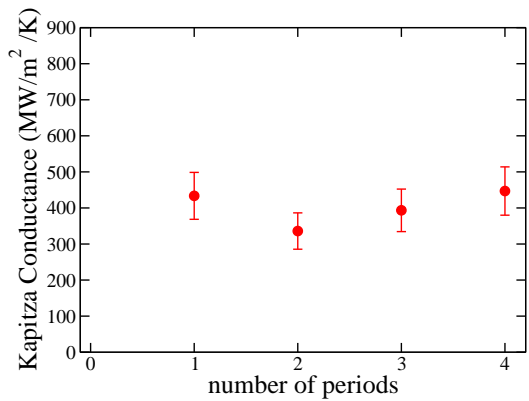}
\caption{(Color online) Thermal boundary conductance of isosceles triangular shaped interfaces having a roughness height $h=32$ MLs and a Period $40$ $a_0$, as a function of the number of periods of the superlattice. Lines are guides to the eye. 
The temperature is $T=40$ K and the mass ratio is $m_r=2$.}
\label{fig_number_periods}
\end{figure}

We remark also that the conductance is higher at high temperatures. Generally speaking, the thermal boundary conductance is found to increase with temperature, a trend often attributed to the existence of inelastic phonon scattering at the interface~\cite{stevens2007,hopkins2009}. This behavior is consistent with our simulation data.
In the following, we will fix the period to $20$ $a_0$ as it leads to moderate finite size effects as already found for superlattices with planar interfaces~\cite{merabia2012}. Finally, since the system we simulate is akin to a superlattice because of the periodic boundary conditions, it is important to probe the effect of the number of periods on the measured conductance. Figure~\ref{fig_number_periods} quantifies the effect of the number of interfaces on the conductance measured in transient simulations. From this figure, we can conclude that within error bars, the number of interfaces has a mild effect on the conductance that we calculate. This is the behaviour expected, as we probe a quantity characterizing the interface solely, independently on the number of interfaces.

\subsection{Effect of the interfacial roughness \label{results_roughness}}

We will now concentrate on the influence of the interfacial roughness on the thermal boundary conductance. 
We will consider rough interfaces having an angle $\alpha$ fixed at $\alpha=45 \deg$, while the height $h$ of the interface is increased so as to change the interfacial roughness, as seen in fig~\ref{schematic_parameters} b. For the following, it is important to keep in mind that, when varying the height of the interface $h$ at a constant value of the angle, the true interfacial area remains constant, and larger by a factor $1/\cos \alpha=\sqrt{2}$ than its corresponding projection on the horizontal xy plane.

Figure~5 displays the evolution of the measured Kapitza conductance, as a function of the interfacial roughness for two temperatures.
 The conductance of a planar interface, which corresponds to the value $h=0$ has been also indicated for the sake of comparison. Two regimes are to be distinguished, depending on the roughness of the interface $h$. When the height is smaller than typically $20$ monolayers, the conductance seems to be constant, or slightly decreases with the roughness, taking values close to the planar interface conductance. When the interfacial height becomes larger, the conductance suddenly increases and tends to saturate for very rough surfaces. 

\begin{figure}
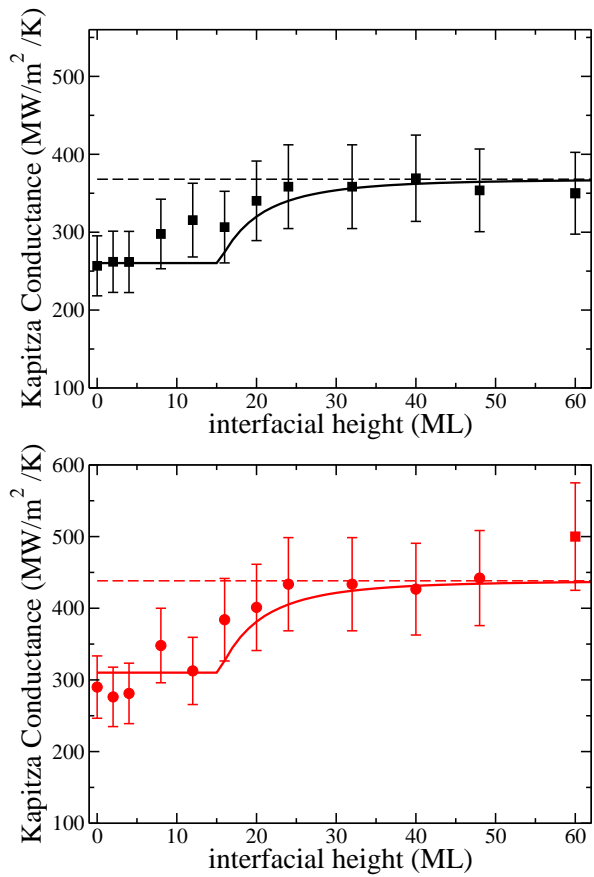

\includegraphics[width=0.9\linewidth]{transient_conductance_energy_ML_2.eps}
\includegraphics[width=0.9\linewidth]{transient_conductance_energy_ML_3.eps}
\caption{(Color online) Thermal boundary conductance of isosceles triangular shaped interfaces, as a function of the interfacial roughness. Top: $T=40$ K; Bottom: $T=18$ K. We have also indicated the conductance of the corresponding planar interfaces 
(ML=0) and very rough interface (ML=60).  
The horizontal dashed lines show the conductance obtained after rescaling the conductance of the planar interface by the true interfacial area. The solid lines denote the theoretical model eq.~(\ref{conductance_model}), with the parameter $\xi=0.2$. The parameters are: total Length=$40$ $a_0$; mass ratio $m_r=2$.}
\label{45deg}
\end{figure}

Interestingly, the increase of the conductance between planar and very rough surfaces is found to be close to the increase of the interfacial area. This is materialized in fig.~5, where we have shown with dashed lines the value of the conductance obtained by multiplying the conductance of a planar interface by a factor $1/\cos \alpha$. 
Depending on the superlattice period, the increase of the Cross-Plane thermal conductivity has been found to be included between $1.3$ and $1.5$, which encompasses the value $\sqrt{2} \simeq 1.41$. This reinforces the message according to which the thermal conductivity of a superlattice is mainly controlled by the Kapitza resistance exhibited by the interfaces, which in turn seems to be primarily governed by the interfacial area. 
 
\begin{figure}
\includegraphics[width=0.7\linewidth]{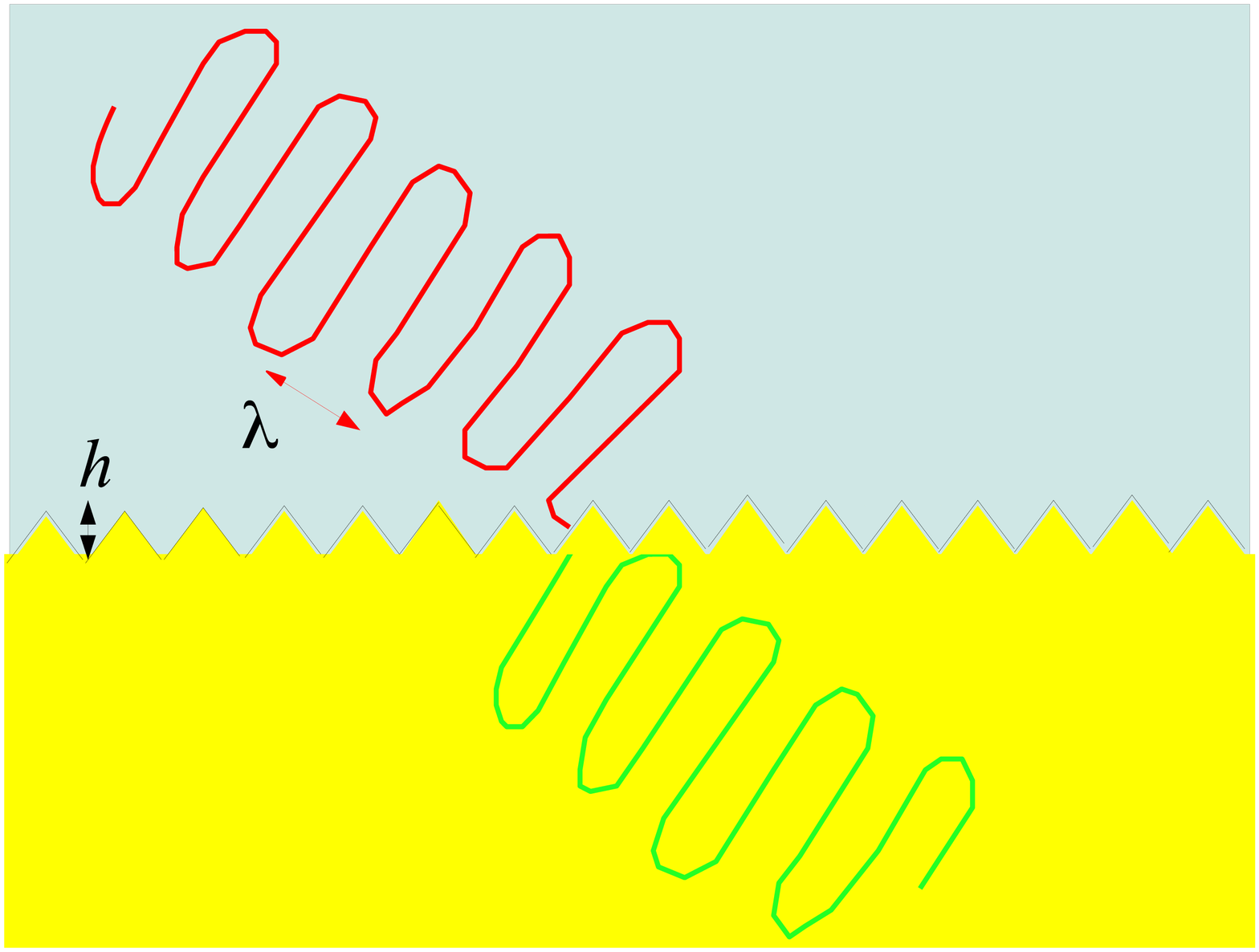}
\includegraphics[width=0.7\linewidth]{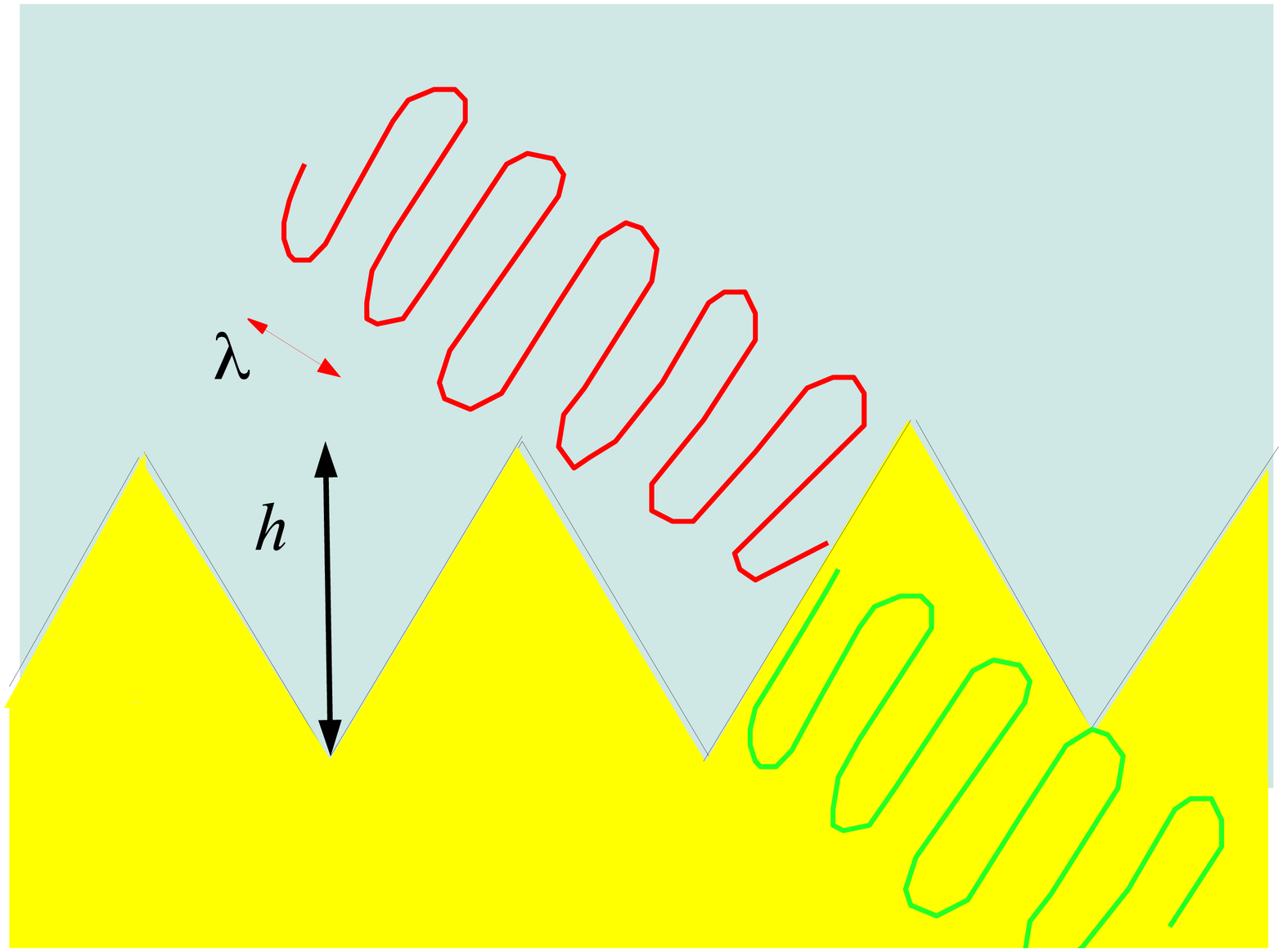}
\caption{(Color online) Schematic representation of the roughness induced phonon scattering. Top: case of a small roughness. The huge majority of incoming phonons 
see the interface as a plane, and the transmitted heat flux is proportional to the projected interface area. Bottom: case of a large roughness. Most of the phonons have a wavelength larger than the interfacial roughness, and the transmitted heat flux is proportional to the true surface area. For the sake of the representation, we have not drawn the reflected waves. Note also that the phonon wavelength is generally not conserved at the passage of the interface.  }
\label{schematic_phonon_roughness}
\end{figure}

We now give some qualitative elements to understand the previous simulation results, regarding the influence of the interfacial roughness on the Kapitza conductance. At this point, it is important to have in mind that in the situations that we have modeled, the energy carriers are phonons which are classically populated. A given phonon mode is characterized among others by its wavelength $\lambda$, which may take practically any value between an atomic distance $2a$ and the simulation box length $L$~\cite{note_phonon_wavelength}.
First, let us concentrate on the case of a small roughness $h$, as exemplified in fig.~\ref{schematic_phonon_roughness}. In this case, the majority of phonon modes have a wavelength larger than $h$, and they see the interface as a planar one: the transmitted heat flux is then controlled by the projected area.
On the other hand, when the interface is very rough, most of the phonons have a wavelength smaller than the height $h$, obviously the phonons no longer feel the interface as planar, phonon scattering becomes completely incoherent, and the transmitted heat flux is controlled by the true surface area. 

To put these arguments on quantitative grounds, we will consider the following expression of the thermal conductance, inspired by the classical AMM model~\cite{little1959}. We introduce a mode-dependent fraction $\psi(\lambda)$, which depends on the considered wavelength, and which is equal to $1$ when the wavelength is supposed to be small compared with the interfacial roughness $h$, and equal $0$ in the opposite case. We define a dimensionless parameter $\xi$, such that:
\begin{eqnarray}
\label{psi}
\psi(\lambda) &=& 1 \; {\rm if} \; \lambda < \xi h \nonumber \\ 
\psi(\lambda) &=& 0 \; \; {\rm otherwise} 
\end{eqnarray}
The parameter $\xi$ will be the adjustable fitting parameter of the model. The interfacial conductance is then supposed to be given by~:
\begin{eqnarray}
\label{conductance_model}
G_K & =& \frac{3}{2} \zeta \rho k_B c_1 ( \int_{0}^{\omega_{\rm Dmin}} g(\omega) \psi(\lambda) d \omega \mathcal I_{12} \nonumber \\
&+ & \frac{\mathcal A}{\mathcal A_0} \int_{0}^{\omega_{\rm Dmin}} g(\omega) (1-\psi(\lambda)) d \omega \mathcal I_{12} )
\end{eqnarray}
where $\rho$ is the crystal number density, $c_1$ is the average sound velocity in medium $1$, $\omega_{\rm Dmin}$ is the Debye frequency of the softer solid. The parameter $\zeta$ is a scaling factor which accounts for the tendency of the AMM model to overpredict the measured Kapitza conductance~\cite{merabia2012}.
The integral $\mathcal I_{12}$ involves the angular dependent transmission coefficient:
\begin{equation}
\label{transmission_coefficient}
\mathcal I_{12}=\int_{0}^{1} \frac{4 Z_1 \mu_1  Z_2 \mu_2}{(Z_1 \mu_1 + Z_2 \mu_2)^2} \mu_1 d\mu_1
\end{equation}
where $Z_i=\rho_i^m c_i$ are the acoustic impedances of the two solids, $\rho_i^m$ being the mass density, and $\mu_1=\cos \theta_1$ is a shorthand notation to denote the cosine of the phonon incident angle~\cite{merabia2012}. Finally, the quantity $\mathcal A/\mathcal A_0$ is the ratio of the true interfacial area over the projected one. The physical motivation of eq.~(\ref{conductance_model}) is simple: phonons having a wavelength $\lambda$ larger than $\xi h$ contribute to a transmitted heat flux proportional to the projected area $\mathcal A_0$, while phonons modes having a wavelength smaller than $\xi h$ contribute to the transmitted heat flux proportionally to the true surface area. We have compared the prediction of eq.~\ref{conductance_model} to the simulation results discussed before. To this end, we have assumed Debye solids, with a vibrational density of states $g(\omega)=\omega^2/(2\pi^2 c_1^3)$, and for the sake of consistency, the mode-dependent wavelength $\lambda$ has been taken to be simply related to the frequency $\omega$~: 
$\lambda=2 \pi c_1/\omega$. Figure~5 compares the predictions of eq.~\ref{conductance_model} to the simulations results. The values of the planar interface conductance has been rescaled by a factor $\zeta=3$ and $4$ at the temperatures $T=18$ K and $T=40$ K respectively. These correction factors account for the fact that the simple AMM model relies on several assumptions-Debye solids-elastic scattering-which may lead to a discrepancy with the MD value. We have chosen the two factors because they yield good agreement with the MD value for smooth interfaces.
 Apart from this rescaling, the parameter $\xi$ has been treated as the only fitting parameter. Figure~5 shows that a good agreement is found using the value $\xi=0.2$, for the two temperatures considered. The smallness of the fitting coefficient may be understood in the following way: consider a given phonon mode: if its wavelength is larger than the roughness $h$, the effective scattering area would be the projected one. On increasing the roughness, $h$ will become comparable with $\lambda$, and the interface will strongly scatter the considered phonon in all directions. This will contribute to slightly diminish the conductance, as compared with the planar case, in agreement with the simulation data points. It is only when the roughness becomes very large as compared with the wavelength $\lambda \ll h$, that interfacial scattering becomes again negligible and the transmitted energy is proportional to the true area. The fitting procedure concludes that this regime is reached when the wavelength becomes smaller than typically one fifth of the interfacial roughness.

\subsection{Effect of the angle of the asperities \label{results_angle}}
\begin{figure}
\includegraphics[width=0.9\linewidth]{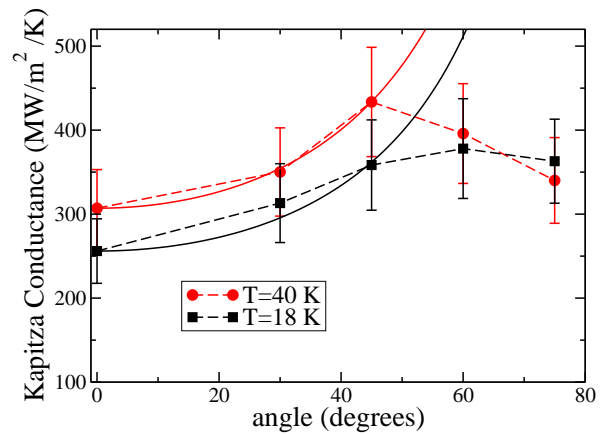}
\caption{(Color online) Thermal boundary conductance as a function of the angle of the asperities $\alpha$. The height of the asperities is fixed here and equal to $24$ ML. The points corresponding to $\alpha=0$ denote the conductance of a planar interface. The solid lines show the interfacial conductance rescaled by the true surface area: $G_K=G_K(\alpha=0)/\cos \alpha$. The other parameters are: Total length=$40$ $a_0$; mass ratio $m_r=2$.}
\label{conductance_angle_fixed_height}
\end{figure}

\begin{figure}
\includegraphics[width=0.9\linewidth]{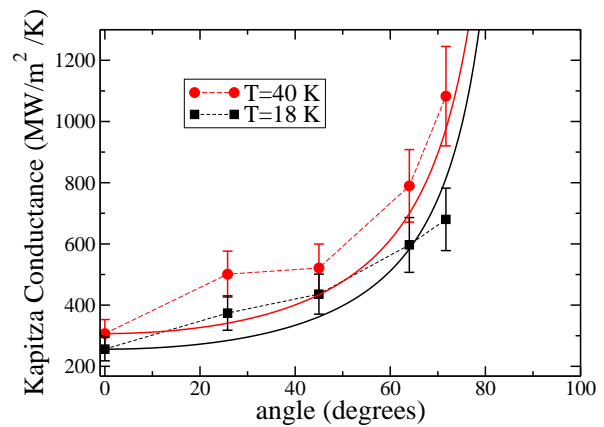}
\caption{(Color online) Same as fig.~\ref{conductance_angle_fixed_height}, but for a constant value of the true surface area. 
The interfacial heights are respectively $h = 21, 34, 43$ and $45$ monolayers for the asperities angles $\alpha=25.8, 45, 64$ and $71.7 \deg$. The solid lines show the interfacial conductance rescaled by the true surface area: $G_K=G_K(\alpha=0)/\cos \alpha$. }
\label{conductance_angle_fixed_surface}
\end{figure}
So far, we have considered the case where the angle $\alpha$ was constant. We now discuss the effect of varying the slope of the model interfaces on the interfacial energy transfer. First, we will change the angle at a fixed value of the interfacial height $h$, as represented in fig.~\ref{schematic_parameters} c. Figure~(\ref{conductance_angle_fixed_height}) shows the evolution of the Kapitza conductance as a function of the angle, at the two temperatures considered. The constant height $h$ retained here corresponds to the regime of large roughnesses in terms of the figure~5 discussed before. We have also indicated the conductance of a planar interface, for the sake of comparison. The evolution of the conductance with the asperities angle seems to be non-monotonous: first, it increases for low angles, reaches a maximum for an asperities angle between $30$ and $45$ degrees, and then decreases when the angle becomes large. In particular, the conductance for an angle greater than $60 \deg$ becomes smaller than the planar interface conductance. This is all the more 
remarkable as in this latter case, the true surface area may increase by a factor $4$ as compared to the planar interface. This discrepancy is best shown after comparing the simulation results with the rescaled conductance $G_K(\alpha=0)/\cos \alpha$, which accounts for the increased surface area induced by the asperities. It is immediately clear, that for the lowest values of the asperities angles, $\alpha=30$ and $45 \deg$, the rescaled conductance seems to describe reasonably interfacial energy transfer. On the other hand, at large values of $\alpha$, the theoretical expression overestimates greatly interfacial transport. Two phenomena may explain the poor conductance reported: first, on increasing the angle, phonon multiple scattering and back-scattering may contribute to diminish interfacial transmission. This has been evidenced by Rajabpour et {\em al.} using Monte-Carlo ray tracing calculations~\cite{rajabpour2011}. Secondly, for the steep slopes interfaces considered here, the effective surface area seems to be the projected one, not the true area, even if the height of the asperities is large. This may be understood qualitatively because for steep interfaces, even if the height is large, the lateral correlation length $l=h/ \tan \alpha$ may become comparable with the phonon wavelengths, and the effective interfacial area becomes the projected one. For these steep interfaces, the regime where the transmitted heat flux is controlled by the true surface area should occur at a very large value of the interfacial height $h$. To verify this assessment, we have run simulations where the true surface area has been kept constant (cf fig.~\ref{schematic_parameters} d). The results are displayed in fig.~\ref{conductance_angle_fixed_surface}, which concludes a different scenario as compared to the evolution shown previously in fig.~(\ref{conductance_angle_fixed_height}). The evolution of the conductance with the angle is no longer non-monotonous as previously observed, but it rather increases monotonously with $\alpha$. For the relatively small values of the angles $\alpha$, the conductance measured may even exceed the rescaled one. We have no interpretation for these large values reported here. Increasing 
further the angle $\alpha$, the simulation data takes values close to the scaled conductances $G_K(\alpha=0)/\cos \alpha$. Note in particular, that the increase of the conductance is pretty large, overpassing the conductance of a planar interface by a factor larger than $3$. In this regime, and for these steep interfaces, it is highly probable that the regime of rough interfaces, in the terms of the previous discussion has been reached: heat transmission becomes controlled by the true surface area. These large enhancement of the Kapitza conductance open the way to design interfaces with tailored interfacial energy transport properties.

\begin{figure}
\includegraphics[width=0.9\linewidth]{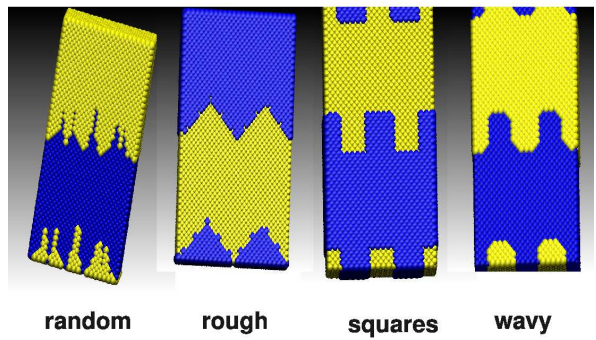}
\caption{(Color online) Illustration of the different interfacial shapes simulated, namely random, rough, square and wave-like 
interfaces.}
\label{illustration_shape}
\end{figure}

\subsection{Effect of the shape of the interfaces \label{results_shape}}
We end up the presentation of the results by discussing the effect of the shape of the interface on the boundary conductance. All the previous discussion concentrated on model triangular interfaces, and it is worth asking how general are the conclusions drawn from the study of the particular type of surfaces. To appraise this question, we have considered different shapes of the interfaces, as depicted in fig.~\ref{illustration_shape}. The common characteristics of these surfaces is the mean interfacial height, here fixed at a value $h=12$ MLs. 
Different morphologies have been designed, ranging from the random interface, to the case of square-like surface and wavy interface obtained by a sinusoidal modulation of the interfacial height.  

\begin{figure}
\includegraphics[width=0.9\linewidth]{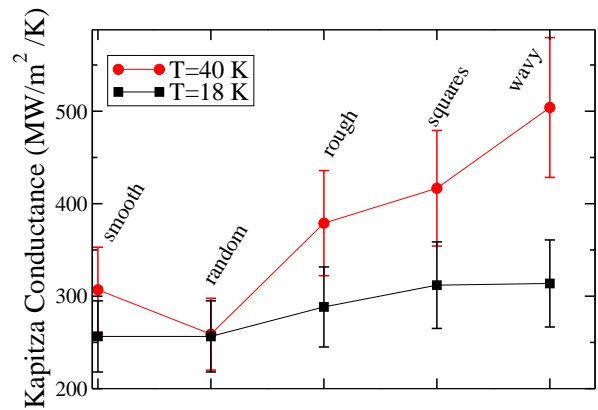}
\caption{(Color online) Thermal boundary conductance for the different interfacial shapes represented in fig.~\ref{illustration_shape}. The height of the different interfaces is fixed here and equal to $12$ ML. The other parameters are: Total Length=$40$ $a_0$; mass ratio $m_r=2$.}
\label{conductance_shape}
\end{figure}

Figure~\ref{conductance_shape} compares the interfacial conductance for the different shapes shown before, at two different temperatures. The relatively large values reported at the highest temperature may be explained by inelastic phonon scattering
taking place between two interfaces.
The shape of the interface seems to affect considerably interfacial transport: random interfaces display a conductance practically equal to the planar interface. Rough interfaces may transfer energy slightly better than planar interfaces depending on the temperature. On the other hand, wavy and square-like interfaces tend to favor energy transmission, the wavy pattern displaying the highest conductance among the different shapes analyzed. These results may be interpreted qualitatively: random and rough shaped interfaces display a distribution of length scales, which tend to promote phonon scattering: even if the global height $h$ is large, in terms of triangular shaped interface, the effective area for the phonons is not the true surface area, but rather the projected area, because $h$ is not the only relevant roughness parameter, and the interaction between incident phonons and small length scale asperities tend to diminish the effective transmission area. On the other hand, regular shaped pattern do not display such a distribution of length scales, and interfacial heat transport becomes controlled by the true surface area: as soon as the majority of phonon modes has a wavelength greater than the single length $h$ characterizing the interfacial morphology, one enters in a "large roughness" regime where the energy transport becomes governed by the true surface area, and the conductance is increased as compared with the planar case. The interfacial conductance is found to be the highest for the wavy interface, because it has the greatest surface area.

\section{Conclusion}
In summary, we have concentrated on the role of the interfacial roughness on energy transmission between solid dielectrics. Thanks to the versatility of molecular dynamics simulations technique, we have conceived model rough surfaces, and probe their ability to conduct heat. The scenario emerging from the simulations is the following: when the roughness introduced is small, most of the phonons see the interface as a planar one and the effective surface area contributing to the transmitted heat flux is the projected area, not the true one. In this regime, one does not expect a Kapitza conductance much different from the planar interface. On the other hand, when the roughness becomes large enough, typically $20$ monolayers in our case, most of the phonons propagating towards the interface are incoherently scattered, and the effective surface area becomes the true surface area. This latter may differ significantly from the projected one, and this is the reason why the boundary conductance of rough interfaces may be greatly enhanced, as compared to planar interfaces. This has 
been demonstrated in this work, with the example of triangular shaped interfaces displaying steep slopes: provided the lateral dimensions characterizing the interfacial roughness is large enough, the increase of the conductance may be threefold. On the other hand, we have probed the conductance of randomly rough interfaces, and shown that they display in general conductances comparable or smaller than atomic planar interfaces. This difference of behavior is explained by the distribution of length scales displayed by the randomly rough surfaces, in comparison with our model patterned surfaces.

The roughness analyzed in this article was large compared to the lattice constants. The case of atomic roughness has been 
more widely addressed in the literature, and wave-packet simulations~\cite{sun2010} give a clear picture of the effect of small atomic roughness on phonon transmission: long wavelength phonons see the interface as ideal and do not contribute to the change of the thermal boundary conductance. On the other hand, short wavelength phonons strongly interact with the small scale roughness, and the corresponding change in phonon transmission is found to depend on the structure of the interface: for regular shaped patterned interfaces, constructive wave interference lead to enhanced transmission thereby increasing the boundary conductance~\cite{tian2012}. Random atomic roughness promotes incoherent phonon scattering, reducing the thermal conductance. These observations are consistent with MD results concerning the cross-plane conductivity of superlattices with rough interfaces: for regularly patterned interfaces, the cross-plane conductivity is slightly greater than ideal superlattices~\cite{termentzidis2011_2}, and the boundary conductance is enhanced~\cite{zhou2013}. When the roughness is random, the cross-plane conductivity shows a small decrease as compared with planar interfaces~\cite{daly2002,termentzidis2011}. The small amplitude of the reduction is related to the small proportion of energy carriers affected by the atomic roughness. Small reductions have been also reported for Si/Ge superlattices with a one layer of interfacial mixing in the incoherent regime of transport~\cite{landry2009b}.  If it is reasonable to rationalize such variations in terms of an atomic interfacial roughness, it is less clear for superlattices having thicker mixed layer. Large enhancements have been observed in this latter case, using MD\cite{stevens2007}. Further work is clearly needed to understand if part of these enhancements is explained by the large scale interfacial roughness.~\cite{tian2012}

Most of the results reported here concern regular shaped patterned interfaces. MD results seem to conclude that these patterned interfaces are good candidates to enhance the intrinsic boundary conductance between two semi-conductors. On the other hand, randomly rough surfaces should be considered if one prefers to reduce the Kapitza conductance between two solids\cite{hopkins2010}. In particular, in the context of superlattices, randomly rough interfaces should be designed if one aims at tailoring materials with the lowest cross-plane thermal conductivity. 

We have also introduced a simple model to rationalize the variations of the thermal boundary conductance as a function of the interfacial height of our model rough interfaces. Further analytical work is clearly desired to understand the interplay between the interface morphology and energy interfacial transport. This will enable to define new directions for the design of interfaces with optimized energy transport properties, with a relative low cost.

\begin{acknowledgments}
Simulations have been run at the "Pole Scientifique de Mod\'elisation Num\'erique" de Lyon using the LAMMPS open source package~\cite{plimpton1995}. We acknowledge interesting discussions with P. Chantrenne, T. Biben, P.-O. Chapuis and D. Lacroix.
\end{acknowledgments}

\end{document}